\documentclass{article}
\usepackage{spconf,amsmath,epsfig}
\usepackage{subfigure}

\title{On Using Multiple Quality Link Metrics with Destination Sequenced Distance Vector Protocol for Wireless Multi-hop Networks}
\name{N. Javaid$^{\ddag}$, A. Bibi$^{\ddag}$, Z. A. Khan$^{\$}$, K. Djouani$^{\dag,\sharp}$}
\address{ $^{\ddag}$Department of Electrical Engineering, COMSATS, Islamabad, Pakistan. \\
        $^{\$}$Faculty of Engineering, Dalhousie University, Halifax, Canada.\\
        $^{\dag}$F'SATIE, TUT, Pretoria, South Africa.\\
        $^{\sharp}$LISSI, Universit\'e Paris-Est Cr\'eteil (UPEC), France.}
\begin{document}
%
\maketitle
\begin{abstract}
In this paper, we compare and analyze performance of five quality link metrics for Wireless Multi-hop Networks (WMhNs). The metrics are based on loss probability measurements; ETX, ETT, InvETX, ML and MD, in a distance vector routing protocol; DSDV. Among these selected metrics, we have implemented ML, MD, InvETX and ETT in DSDV which are previously implemented with different protocols; ML, MD, InvETX  are implemented with OLSR, while ETT is implemented in MR-LQSR. For our comparison, we have selected Throughput, Normalized Routing Load (NRL) and End-to-End Delay (E2ED) as performance parameters. Finally, we deduce that InvETX due to low computational burden and link asymmetry measurement outperforms among all metrics.
\end{abstract}

\vspace{-0.3cm}
\section{Introduction}
\vspace{-0.3cm}
A routing protocol is responsible for significant performance from the underlying wireless network. A routing link metric is a key component of a routing protocol. As, it finds all possible end-to-end routes and also the fastest route. Minimum Hop-count; non-quality link metric is the most popular and is IETF standard metric [1]. It is appropriately used by Wireless Ad-hoc Networks, where the objective is to find new paths as fast as possible in the situations where quality paths cannot be found quickly and/or can not efficiently work because of higher rates of node mobility. Moreover, hop-count is the simplest to calculate and it avoids any computational burden on the routing protocol. It is obvious from its equation: $Hop\_Count_{P_{e2e}}=\displaystyle\sum_{l \in P_{e2e}}l$.

Quality Link Metrics (QLMs) are firstly introduced in [2], for successful delivery of data packets in static networks. Efficiency of static WMhNs depends upon low routing latency, minimized routing load, and less end-to-end delay (E2ED). To achieve proficient performance of a protocol in such networks a realistic QLM is needed.

Several QLMs have been proposed, like, Expected Transmission Count (ETX)[2], Expected Transmission Time (ETT) [3], Interference and Bandwidth Adjusted ETX (IBETX) [4], Expected Link Performance (ELP) [5], Minimum Loss (ML) [6], Minimum Delay (MD)[7] and Inverse ETX (InvETX) [8]. The metrics, ETX, ML, MD and InvETX have already been implemented [8] with a proactive routing protocol, Optimized Link State Routing (OLSR) [9] using Link State routing technique. While, ETX, IBETX, and ELP are implemented with Destination Sequenced Distance Vector (DSDV) [10] based on distance vector routing algorithm. Distance vector routing uses the next hop information during exchanging the routing information, whereas link state information contains the whole topological information. In this paper, we implement ETX, ML, MD, ETT and InvETX in distance vector protocol DSDV. Original ETX is implemented with DSDV[2].

In this paper, we have implemented five QLMs; ETX, InvETX, ETT, ML and MD with DSDV which is based on distance vector algorithm. Previouse implementation of QLMs in routing protocols have not considered routing load for performance measurements. We, in our previous work [8] have considered routing load while analyzing the performance of OLSR which is a link state based proactive routing protocol. Now, in the same way, we are evaluating the performance of those link metrics along with ETT in this paper. Moreover, ML and MD are implemented in OLSR, and ETT is implemented with MR-LQSR, on the other hand, we implement these metrics along with InvETX in DSDV.

\vspace{-0.3cm}
\section{Quality Link Metrics}
\vspace{-0.3cm}
[2] is the very first work launching the idea of quality routing by proposing ETX. In this section, we discuss five QLMs, among all are based on ETX except MD. A detailed study on ETX-based metrics can be found in [1].

\textbf{(1) ETX:} Forward and reverse loss rates and link asymmetries of the links are measured by calculating the loss probabilities of links by broadcasting probe packets. In this approach, each node is supposed to periodically send out a broadcast probe packets only to the neighbors without any retransmission. Nodes track the number of successfully received probes from each neighbor during a sliding window time; 10 seconds, and include this information in their own probes. Nodes can calculate reverse loss probability; $d_r$, directly from the number of probes they receive from a neighbor in the time window, and they can use the information about themselves received in the last probe from a neighbor to calculate forward loss probability;$d_f$.

\vspace{-0.5cm}
\small
\begin{eqnarray}
 ETX_{P_{e2e}}=\sum_{l \in P_{e2e}}^{}\frac{1}{(d_f^{(l)}\times d_r^{(l)})}
\end{eqnarray}
\normalsize

\textbf{(2) InvETX:} remarkably avoids the computational overhead and thus achieves least delay [8]. ETX calculates the inverse of probability of success (product of forward and reverse probabilities) but as the names implies, InvETX directly computes probabilities.

\vspace{-0.5cm}
\small
\begin{eqnarray}
 InvETX_{P_{e2e}}=\sum_{l \in P_{e2e}}^{}{(d_f^{(l)}\times d_r^{(l)})}
\end{eqnarray}
\normalsize

\textbf{(3) ETT:} of a link as a "bandwidth-adjusted ETX" is defined in [3]. Authors consider the link bandwidth to obtain the time spent in transmitting a packet. They start with ETX and divide by with link bandwidth. Let $S$ denote the size of the packet and $B$ the bandwidth (raw data rate) of the link $l$. Then:

\vspace{-0.5cm}
\small
\begin{eqnarray}
ETT_l=ETX_l\times t_l
\end{eqnarray}
\normalsize

\vspace{-0.5cm}
\small
\begin{eqnarray}
ETT_l=ETX_l\times\frac{S_F}{B_l}
\end{eqnarray}
\normalsize

\vspace{-0.5cm}
\small
\begin{eqnarray}
ETT_l=ETX_l\times(\frac{S_F}{\frac{S_L}{T_S-T_L}})_l
\end{eqnarray}
\normalsize

\vspace{-0.5cm}
\small
\begin{eqnarray}
ETT_{P_{e2e}}=\sum_{l \in P_{e2e}}^{}ETX_l\times(\frac{S_F}{\frac{S_L}{T_S-T_L}})_l
\end{eqnarray}
\normalsize

Forward and reverse loss rates of links is measured in ETX portion of ETT. These lost rates are calculated through broadcast prob packet as in [2]. The problem of determining the bandwidth of each link is more complex. For the measurement of bandwidth, ETT uses the technique of packet pairs, i.e, after every minute, each node is supposed to send two back-to-back probe packets to each of its neighbors. First probe of size of 137 bytes, while the second probe packet is comparatively heavier and is of 1137 bytes. Upon receiving these two probes, neighbor measures the time difference between the reception of the first and the second probe and acknowledges the sender with the difference. To estimate the bandwidth, sender takes the minimum of 10 consecutive samples and then divides the size of the second probe packet by the minimum size sample.

\textbf{(4) ML:} is based on ETX with the aim of selecting the path with the minimum loss probability. It uses the probability of successful transmissions, and not the inverse probability, as in ETX. Another difference of ML with ETX is that ETX finalizes the end-to-end route with two or three links, whereas, like InvETX and IBETX, ML also considers the longer paths. The whole route's probability is given by the product of the links' probabilities instead of the sum of their inverse probabilities (like ETX). It has the advantage of eliminating the routes with high loss rate, and the disadvantage that some low quality links may still be taken into account in choosing a given route, since the metric considers only the total probability product [11].

\vspace{-0.5cm}
\small
\begin{eqnarray}
ML_{P_{e2e}}=\prod_{l \in P_{e2e}}^{}(d_f^{(l)}\times d_r^{(l)})
\end{eqnarray}
\normalsize

\textbf{(5) MD:} With MD metric, routing table computation is based on the total minimum transmission delay. The transmission delay measurements come from a variant of a link capacity estimation technique, known as Ad-hoc Probe. The technique takes into account the differences in clock synchronization, thus providing a more reliable measurement. A disadvantage is that this metric considers routes which have nodes sharing a collision domain with many other nodes, and this tends to degrade the communication on such routes. The Ad-Hoc Probe algorithm uses packet-pairs to measure the packet dispersion [11]. The formula used to calculate packet dispersion $T$ from the packet pair sample is given by:

\vspace{-0.5cm}
\small
\begin{eqnarray}
T=T_{recv2,i}-T_{recv1,i}
\end{eqnarray}
\normalsize


\vspace{-0.5cm}
\small
\begin{eqnarray}
T=(T_{recv2,i}-T_{send,i}-\delta)-(T_{recv1,i}-T_{send,i}-\delta)
\end{eqnarray}
\normalsize

Where $\delta$ is clock offset of nodes, $T_{send,i}$ is packet sending time stamped by sender, whereas $T_{recv1,i}$, and $T_{recv2,i}$ are receiving time of each packet stamped by receiver node.

\vspace{-0.3cm}
\section{Important Issues Regarding QLMs}
\vspace{-0.3cm}
Here, we discuss the direct influences of mathematical design of QLMs on the performance of routing protocol implementing it and indirect affects on efficiency of the underlying network being operated by respective protocol.

\textbf{(1) Link asymmetry:} Link asymmetry can be used by QLMs to check the loss ratios of a link in both directions; forward and backward. If the asymmetry of the link is not determined correctly then route entry is downgraded to unidirectionality questionable. If a route request is received over such a link, the node delays forwarding it while it issues a direct, one-hop unicast route request back to the questionable neighbor. If acknowledgement is received back to the sender, then node forwards the original route request and confiscates the blacklist entry, otherwise, request is dropped by node. In ETX, invETX and ML account link loss ratios, in both directions of link.

\textbf{(2) Low computational overhead:} For routing metric, necessary computations should be considered that must not consume memory, processing capability and the most important; battery power [8]. They discuss the case of three widely used routing link metrics for wireless routing protocols: ETX, invETX, ETT and ML, as in Fig.1.

\textbf{(3) Low routing overhead (routing latency and routing load):} Routing overhead is discussed in detail in [12]. Wireless networks have limited bandwidth Computing a link metric in such a way that it generate extra routing overhead reduce packet delivery.

\begin{figure}[!h]
  \centering
 \subfigure{\includegraphics[height=2 cm,width=4.2 cm]{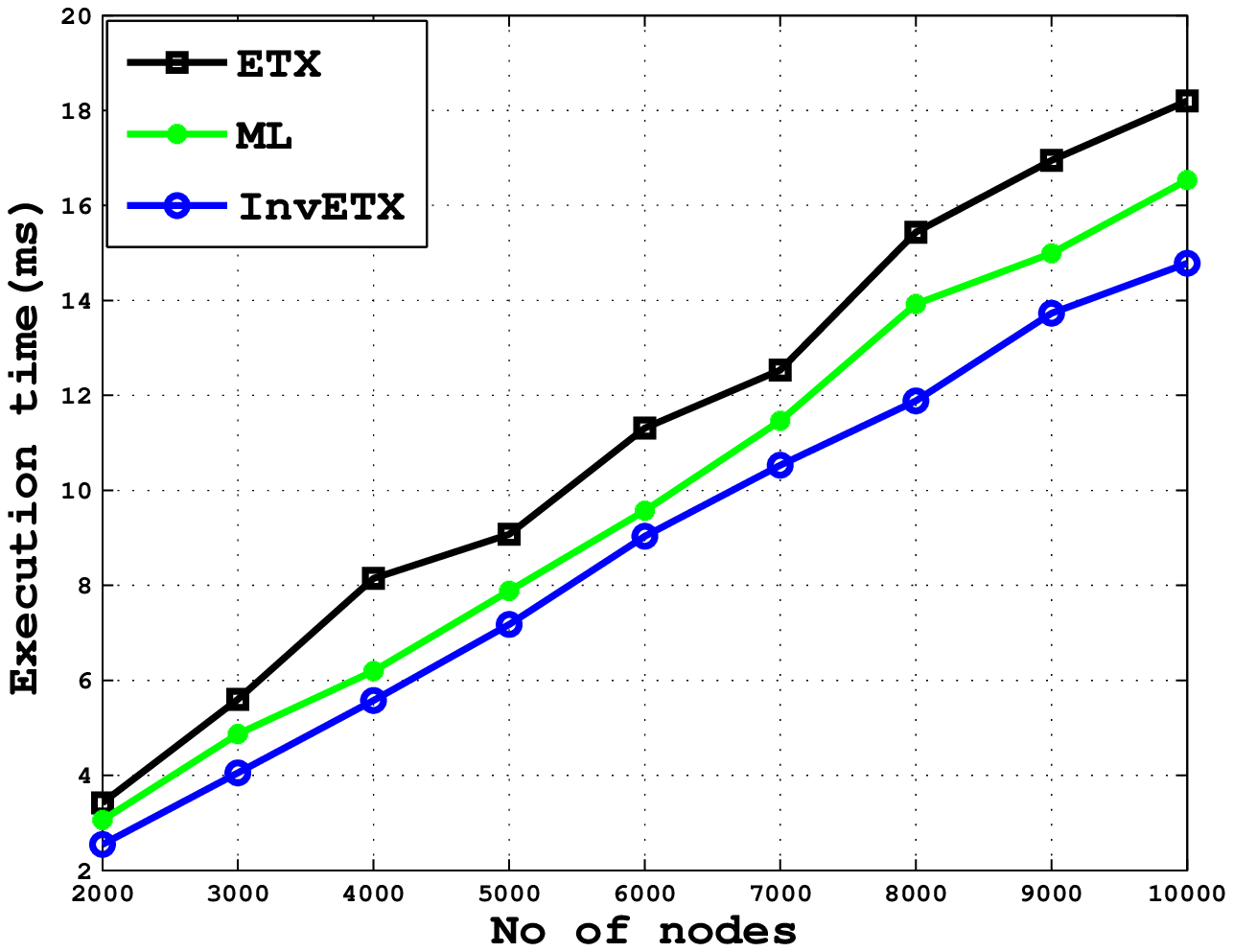}}
 \subfigure{\includegraphics[height=2  cm,width=4.2 cm]{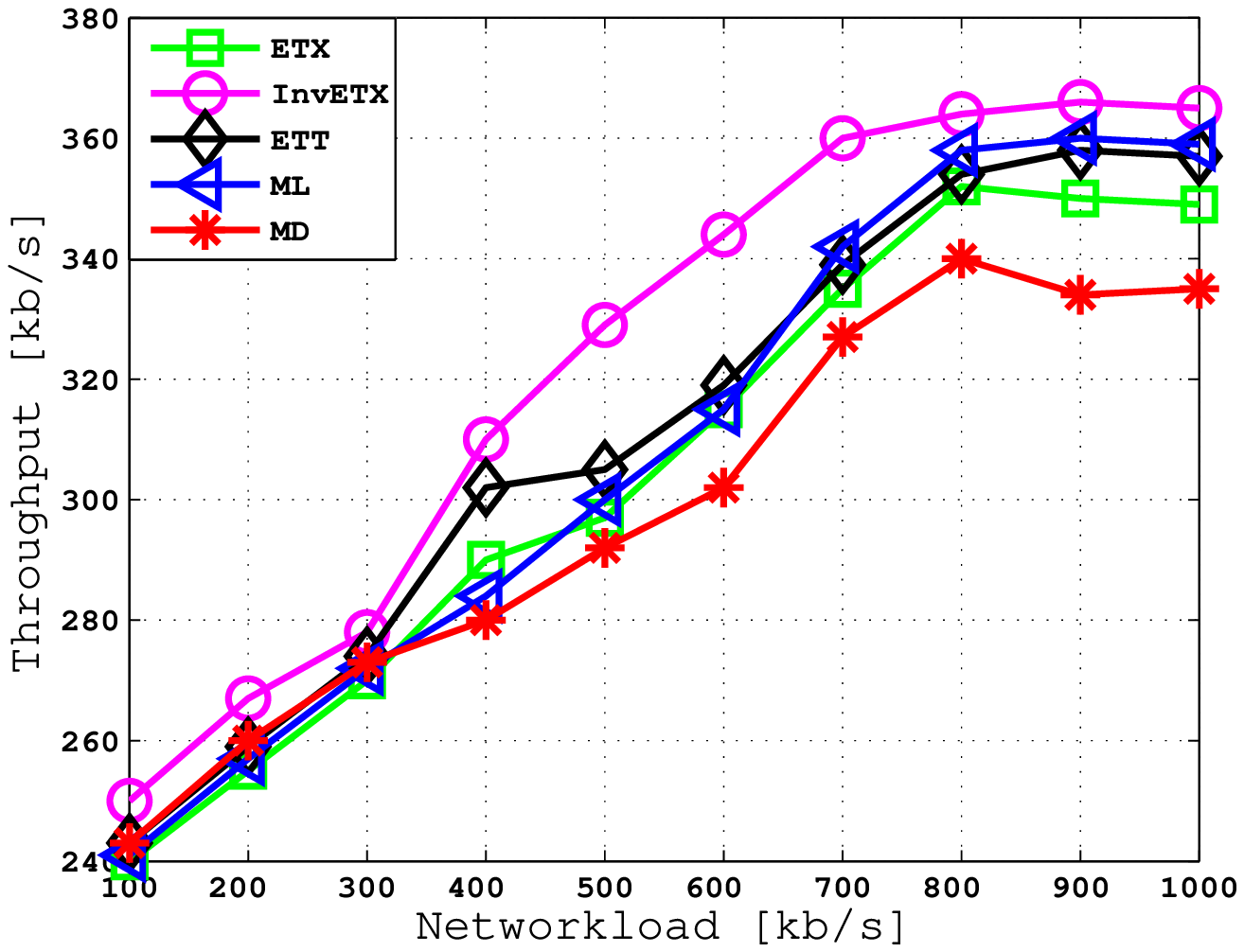}}
  \subfigure{\includegraphics[height=2 cm,width=4.2 cm]{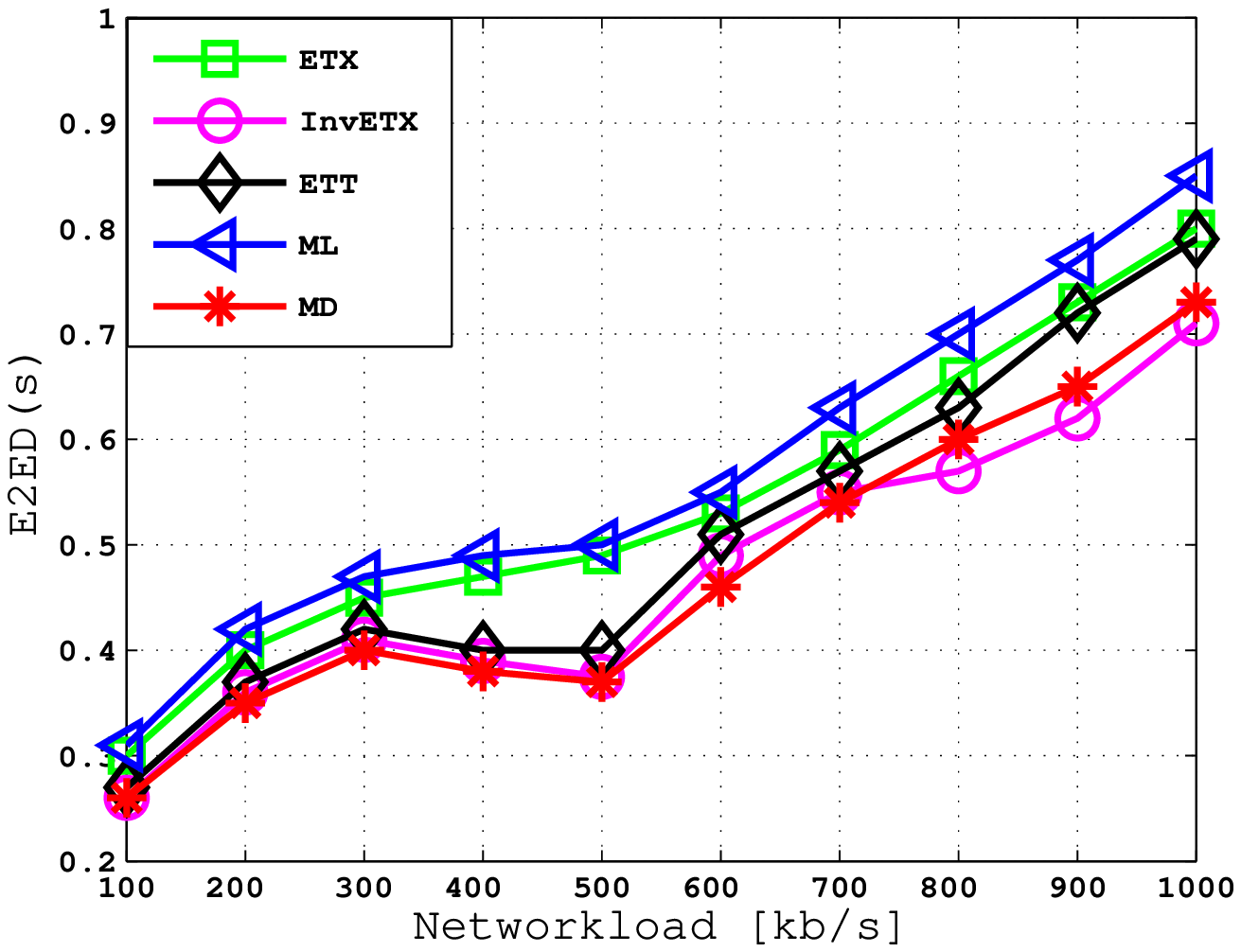}}
 \subfigure{\includegraphics[height=2  cm,width=4.2 cm]{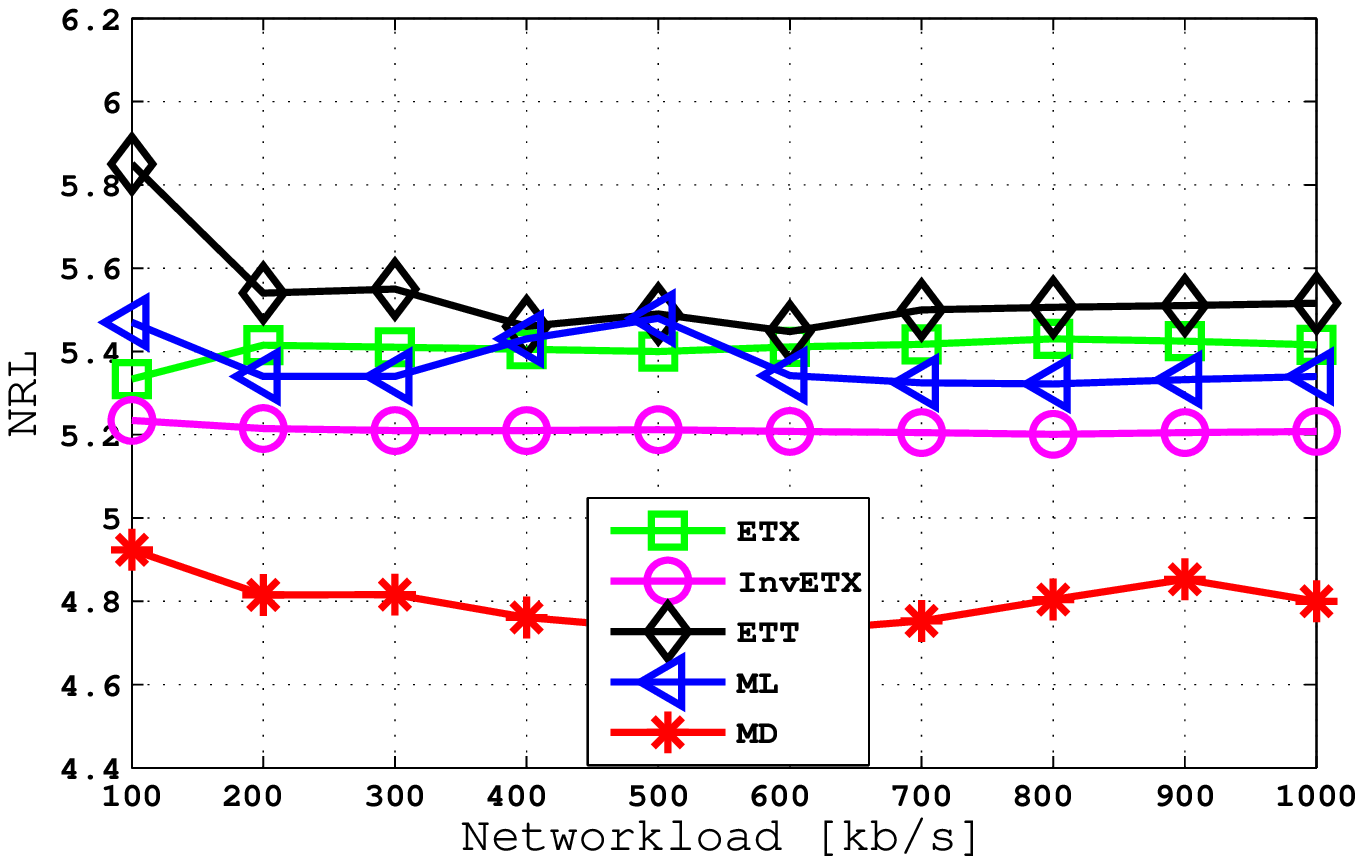}}
  \caption{Simulation Results for Modeled Framework}
\end{figure}

\textbf{(4) Trade-offs:} Generally, a protocol achieves higher throughput values at the cost of increased routing latency in the case of static networks. Whereas, in mobile networks, where link breakage is frequent cause more routing load to obtain better throughput from the network [8]. A QLM in a specific scenario supposed to a suitable trade-off between routing latency and routing load to achieve optimal performance.

\vspace{-0.3cm}
\section{Simulation Setup and Performance Evaluation of QLMs}
\vspace{-0.3cm}
This section provides the details concerning the simulation environment. We compare the performance of ETX, InvETX, ML, ETT and MD in NS-2. The window $w$ used for link probe packets is chosen to be of size 10$s$. The wireless network consists of 50 nodes randomly placed in an area of 1000\textit{m} x 1000\textit{m}. The 20 source-destination pairs are randomly selected to generate Continuous Bit Rate (CBR) traffic with a packet of size 640$bytes$. To examine the performance of QLMs under different network loads, the traffic rate is varied from 1 to 10 packets per second. For each packet rate, the simulations are run for five different topologies for 900$s$ each and then their mean is used to plot the results. Wireless networks suffer from bandwidth and delay. Because of on-demand nature, the reactive protocols are best suited to cope with these issues for mobile scenario where change in topology is frequent. We are dealing with static networks where proactive protocols work at their best because of getting the picture of whole topology and independent of the data generation. Performance of QLMs has been evaluated and then compared with three performance parameters; throughput, E2ED, and NRL.

%
%

\textbf{(1) Throughput:} Among selected five QLMs; ETX (eq.1), InvETX (eq.2), ETT (eq.6), ML (eq.7) and MD (eq.9), MD is not considering link asymmetry. While ETT, ETX, and ML are introducing computational overhead, as depicted in Fig.1. Forward and reverse probes are sent by ETX, ETT, InvETX and ML to check link asymmetry periodically. Asymmetric links and computation overhead can increase drop rate by introducing computational delay and delivering data to unreliable path due to lack of correct assumption about link status. ETX due to computational overhead increases delay, therefore, produces lower throughput value. ETT uses extra probes after every minute by sending two packets of 137 bytes and 1137 bytes to estimate the bandwidth. This not only introduces routing load due to extra probes at routing layer but also computational overhead is introduced by estimating the bandwidth from these probes. In higher network loads, network is more sensitive for routing load and delay. Computational overhead and routing load lead to more drop rate in high data traffic rates.  Lack of link asymmetry in MD produces more drop rates. ML uses product of drop rates for a link as well as for a complete path, and introduces computational overhead (as shown in Fig.1). InvETX produces lowest computational overhead (Fig.1) by taking the product of forward and reverse delivery rates of a link and selecting the maximum InvETX value of a path which is sum of individual links of a path. So, InvETX achieves high throughput, as obvious from Fig.2. Unlike MD, InvETX calculates link asymmetry as well as contains low routing load (Fig.4) by avoiding the use of extra probes after every minute like in ETT.

\textbf{(2) E2ED:} is the time a packet takes to reach the destination from the source. We have measured it as the mean of Round Trip Time (RTT) taken by all packets. Computational delay and longer paths cause latency for end-to-end path calculation. MD estimates the paths based upon one way delay, while ETT selects paths with shorter delay based on the bandwidth estimation. Therefore in medium and low network loads, both metrics produce lower delay (Fig.3), while in high network loads, ETT due to computational overhead and MD due to lack of measuring the link asymmetry augments E2ED. InvETX as compared to ETX and ML has lowest computational overhead thus selecting the paths with low delay, as shown in Fig.3.

\textbf{(3) NRL:} is the number of routing packets transmitted by a routing protocol for a single data packet to be delivered successfully at destination. ETT and MD both are supposed to calculate the paths with low latency. In ETT, bandwidth estimation is considered to select a path with low latency. MD produces the lowest value of routing load (Fig.4) because it generates delivery measurement probes only in forward direction. On the other hand, ETX, InvETX and ML send both forward and reverse delivery probes to check link asymmetry thus producing more routing load as compared to MD. The highest NRL among the selected metrics is produced by ETT, because it uses forward and reverse delivery probes with a pair of packets for calculating low E2E path. Each node is supposed to send two back-to-back probe packets to each of its neighbor after every minute. First probe packet is small and having the size of 137 bytes, while the second probe packet is comparatively heavier and of 1137 bytes. Upon receiving these two probes, neighbor computes the time difference between the reception of the first and the second probes and sends acknowledgement value back to the sender. For bandwidth estimation, sender takes the minimum of 10 consecutive samples and then divides the size of the larger probe packet by the minimum sample value.

\vspace{-0.3cm}
\section{Modeling Routing Overhead by DSDV with Selected QLMs}
\vspace{-0.3cm}
For computing routes by using QLMs, loss probabilities may be required. In OLSR, while computing the loss probabilities, modified HELLO messages are used. Whereas, DSDV is supposed to send extra small probes for measuring the loss probabilities and this leads to more routing load. As in this work, we are considering the routing load as a metric to evaluate the performance; therefore, we first discuss the route maintenance operation of DSDV. DSDV periodically exchanges link state updates with its neighbors to maintain the recent information about connectivity in the network. Moreover, routes are updated thorough trigger updates also. The periodic route updates are flooded with full dump period, while trigger update flooding takes place through incremental dumps only when a link is broken in an active route. Although, the trigger route update operation may appear surplus because of the employment of link state monitoring periodically, it has certain advantages. Monitoring the link status periodically leads to routing loops which are eliminated in trigger route updates using the latest sequence numbers. To keep updated all nodes in a network with topological information, each routing protocol has to exchange routing packets generating routing overhead. For this overhead the protocol has to pay some cost in the form of energy consumed per packet. So, we define this cost in the equation given below to calculate the routing overhead in terms of packet cost. First two costs in eq.10 (eq.10a, 10b) are the same as we have defined in [12]. We define third part (eq.10c) for this work. $C_{E-metric}^{DSDV}$ shows the cost of those packets which are to be sent for measuring a QLM.

\vspace{-0.5cm}
\small
\begin{eqnarray}
C_{E-total}^{DSDV}=C_{E-per}^{DSDV}+C_{E-tri}^{DSDV}+C_{E-metric}^{DSDV}
\end{eqnarray}
\normalsize

Expressions for $C_{E-per}^{DSDV}$ and $C_{E-tri}^{DSDV}$ are same as in [12] and can be further studied from [13].

\vspace{-0.5cm}
\small
\begin{eqnarray}
C_{E-Per}^{(DSDV)}=\int _{0}^{\tau1}(P_{err}d_{avg} + d_{avg} \sum _{i=0}^{h-1}(P_{err})^{i+1}\prod _{j=1}^{i}d_f[j])(10a)\nonumber
\end{eqnarray}
\normalsize

\vspace{-0.5cm}
\tiny
\begin{eqnarray}
C_{E-Tri}^{(DSDV)}=\int _{0}^{\tau2}\sum_{p=1}^{M} \sum_{n=1}^{N} (1-P_{nlb})_n P_{err}d_{avg} + d_{avg} \sum _{i=0}^{h-1}(P_{err})^{i+1}\prod _{j=1}^{i}d_f[j](10b)\nonumber
\end{eqnarray}
\normalsize

\vspace{-0.5cm}
\small
\begin{eqnarray}
C_{E-metric}^{DSDV}=C_{E-QLM}^{ETX,InvETX,ML}, C_{E-QLM}^{ETT}, C_{E-QLM}^{ML}(10c)\nonumber
\end{eqnarray}
\normalsize

\vspace{-0.5cm}
\small
\begin{eqnarray}
C_{E-QLM}^{ETX,InvETX,ML}=(\alpha_{d_f}+\alpha_{d_r})\times \tau_{NL}
\end{eqnarray}
\normalsize

\vspace{-0.5cm}
\tiny
\begin{eqnarray}
C_{E-QLM}^{ETT}=C_{E-QLM}^{ETX,InvETX,ML}+\int_{}^{\tau_{NL}}\alpha_{s-probes}+\alpha_{l-probes}
\end{eqnarray}
\normalsize

\vspace{-0.5cm}
\small
\begin{eqnarray}
C_{E-QLM}^{MD}=2\times\alpha_{d_f}\times\tau_{NL}
\end{eqnarray}
\normalsize
\vspace{-0.5cm}

Where, $\alpha_{d_f}$ and $\alpha_{d_r}$ are the rates of forward and reverse probe deliveries. $\tau_{NL}$ is the total network life time or total simulation time. Similarly, for ETT, $\alpha_{s-probes}$ and $\alpha_{l-probes}$ are the rates of exchange of small and large probes.

\vspace{-0.3cm}
\section{Conclusion}
\vspace{-0.3cm}
Routing protocols are responsible for finding efficient route selection mechanism for reliable communication by selecting optimal routes. There are two types of link metrics; non-quality link metrics and quality link metrics. In this paper, we have compared and analyzed the performance of five quality link metrics which are based on loss probability measurements; ETX, ETT, InvETX, ML and MD. For comparison, we have selected distance vector routing algorithm protocol; DSDV. We implemented ML, MD, InvETX and ETT in DSDV and computed computational burden of loss probability measurements in ETX, InvETX and ML. It is analyzed that ETX and ML produce more computational burden when compared with InvETX. MD does not measure the link asymmetry, thus fails to achieve appreciable throughput for the operating protocol. InvETX due to low computational overhead and accurate link asymmetry measurement outperforms in DSDV among five selected quality link routing metrics.

\end{document}